\documentclass[prb,aps,twocolumn,showpacs]{revtex4}
\usepackage{graphicx}
\usepackage{dcolumn}
\usepackage{bm}
\usepackage{color}
\begin{document}

\title{Stationary phase approach to the quasiparticle interference
on the surface of three dimensional strong topological insulators}
\author{ Qin Liu \footnote{E-mail: liuqin@mail.sim.ac.cn}}
\affiliation{State Key Laboratory of
Functional Materials for Informatics, Shanghai Institute of
Microsystem and Information Technology, CAS, Shanghai 200050, China}

\date{\today}

\begin{abstract}
Constant energy contour (CEC) of the surface bands in topological insulators
varies not only with materials but also at different
energies. The quasiparticle interference caused by scattering-off from
defects on the surface of topological insulators is an effective way to
reveal the topologies of the CEC and can be probed by scanning tunneling
microscopy (STM). Using stationary phase approach,
a general analytic formulation of the
local density of states as well as the power laws of the Friedel oscillation
are present, based only on the time-reversal symmetry and the local geometry around
the scattering end points on the CEC. Distinct response of surface states
to magnetic impurities from that of nonmagnetic impurities is predicted
in particular,
which is proposed to be measured in a closed ``magnetic wall'' setup on the surface of
topological insulators.
\end{abstract}

\maketitle

\section{Introduction}
Topological insulators in three dimensions (3D) are band insulators which have
bulk insulating gap but gapless surface
states with odd number of Dirac cones protected by the time-reversal
symmetry (TRS). As an useful surface probe, recent angle-resolved photoemission spectroscopy (ARPES)
experiments demonstrate clearly that the bismuth-based class of
materials, Bi$_2$X$_3$ (X=Te or Se), are 3D strong topological
insulators (TI) with a single Dirac cone on the
surface \cite{Chen2009,Xia2009,Hsieh2009}. The effective surface Hamiltonian
when close to the Dirac point is given by $H_0=\hbar v_F\hat z\cdot({\bf \sigma\times
k})$, which describes the helical Dirac fermions --- charge carriers
which behave as massless relativistic particles with spin locked to its
momentum. However, compared to the familiar Dirac fermions in particle physics, those
emergent quasiparticles from nontrivial surface states of 3DTIs exhibit richer behaviors.
In Bi$_2$Te$_3$, an unconventional hexagonal warping effect, $\lambda(k_+^3+k_-^3)\sigma_z$, appears
due to the crystal symmetry \cite{Fu2009}, under which the constant energy contour (CEC)
of surface band evolves from a convex circle to a concave hexagon
as the Fermi surface climbing up away from the Dirac point.
This hexagonal warping effect is responsible for the observed snowflake shape Fermi surface \cite{Chen2009},
being geometrically not as simple as a circle, new interesting phenomena are expected to arise.

Quasiparticle interference (QPI) caused by scattering-off from defects on the surface
of 3DTIs is an effective way to reveal the topological nature of the surface states.
The interference between incoming and outgoing waves at momenta ${\bf k}_i$ and ${\bf k}_f$
leads to an amplitude modulation, Friedel oscillation \cite{Friedel1952},
in the local density of state (LDOS) at wave vector
${\bf q}={\bf k}_f-{\bf k}_i$. Nowadays, such modulation can be studied by one more powerful
surface probe, scanning tunneling microscopy (STM), directly in real space and provide
information in momentum space through Fourier transform scanning tunneling spectroscopy (FT-STS).
Several STM measurements \cite{Xue2009,Alpichshev2009,Manoharan2009,Yazdani2009}
with ordinary (spin-unpolarized) tip have been performed on the surface of 3DTIs
in the presence of nonmagnetic point and step impurities, and the following features
share in common. (i) The topological suppression of backward scattering from nonmagnetic point and
edge impurities is confirmed by the observation of strongly damped oscillations in LDOS,
companied with the invisibility of the corresponding scattering wave vector ${\bf q}$ in FT-STS.
(ii) Anomalous oscillations are reported in Bi$_2$Te$_3$
for both point and edge impurities when the hexagon warping effect
starting to work.
(iii) Surface bound states exist when near to the point and edge impurities.
These experimental facts are theoretically well-explained
by different groups \cite{Zhou2009,Lee2009,Guo2010,Biswas},
however no general analytic expressions of the LDOS have been presented yet.
One conclusion in common is now clear that for short-range impurities,
different from the well-known $R^{-1}$ and $R^{-1/2}$ power laws of Friedel oscillations
in two-dimensional electron gas (2DEG) \cite{Crommie1993} for point and edge impurities,
the leading powers of Friedel oscillations in helical liquid are dominated by the
scattering between time-reversal end points (TRP) and are suppressed respectively to
$R^{-2}$ and $R^{-3/2}$. This result is the crucial reason of
the invisibility of the scattering wave vector ${\bf q}$ in FT-STS,
and is the direct consequence of the forbiddance of backscattering protected
by TRS in helical liquid.
Bearing in mind also the reported anomalously pronounced oscillations
in LDOS for
both point and edge impurities \cite{Xue2009,Alpichshev2009,Manoharan2009,Yazdani2009}
at bias voltages where the CEC is noncircular but shapes as a snowflake,
we become to realize that two ingredients are essential
to the oscillations of the LDOS, namely, the TRS of the scattering end points
and the geometry of the CEC.

Motivated by these arguments, in this work,
we present a general analytic formulation of the LDOS measurement in
STM experiments using the stationary phase approach. \cite{Ruth1966}
This approach has been taken successfully to study the Ruderman-Kittel-Kasuya-Yosida
interaction in 3D systems with nonspherical Fermi surfaces  \cite{Ruth1966},
and unusual correlations are obtained.
The advantages of this approach reside in that it is sensitive only to the local geometry
around the so-called ``stationary points'' on the CEC,
and can be applied not only to bulk states but also to surface states.
Using this approach, the followings are obtained.
First, a complete result of the power-expansion series of the measured LDOS is tabled
for both point- and edge-shaped nonmagnetic and magnetic impurities,
modeled by delta potentials, with ordinary and spin-polarized STM tip detection.
These results depend only on the TRS and the local geometry around the scattering end points
on the CEC at the energy of interest, which explain not only the usual
$R^{-1}$ and $R^{-1/2}$ power laws in 2DEG but also the famous $R^{-2}$ and $R^{-3/2}$
oscillations in helical liquid. Second, the most general formulation of the LDOS
from CEC beyond the hexagon warping effects is also present,
which plays the role of a first-glace guide for
future ordinary and spin-polarized STM experiments
on potential TI materials with more complicated Fermi surfaces.
Furthermore, spin-polarized STM experiments are focused on in particular.
It is found that there is no signal for nonmagnetic impurities which is
in consistence with the TRS. While for magnetic impurities, pronounced oscillations
of LDOS are predicted irrelevant of the TRS of the scattering end points on the CEC,
which means the backscattering channels are opened via spin-flip processes \cite{Pan2011}.
A few ARPES experiments with magnetic doping are performed \cite{Chen2010,Pan2011}, whereas
the spin-polarized STM ones are still lacking and called on.
Finally, it is emphasized that our results can
also be generalized to surface states of TIs in higher dimensions \cite{Qi2008}.

We mention that to compare with the recent experiments \cite{Xue2009,Alpichshev2009,Manoharan2009,Yazdani2009}
in detail, some complexities \cite{Zhou2009,Lee2009,Guo2010,Biswas}
should still be involved beyond the delta-potential assumption
when considering the scattering processes, however this is not the main focus of our work.

The rest of this paper is organized as follows. In Sect. \ref{standingwave}, we
discuss the intuitive picture of the interference between
helical waves scattered by magnetic impurities
which break the TRS.
In Sect. \ref{general formulation}, we
present the general analytic formulation of LDOS for point and
edge impurities respectively by focusing firstly on
those CEC where the stationary points are extremal points. We then generalize the
results to the CEC where the order of the first nonzero expansion coefficient
around the stationary points is greater than two.
This work is finally concluded in Sect. \ref{conclusion}.

\section{Standing wave of the spin interference between two helical waves}
\label{standingwave}
\begin{figure}[tbp]
\begin{center}
\includegraphics[width=2.8in]{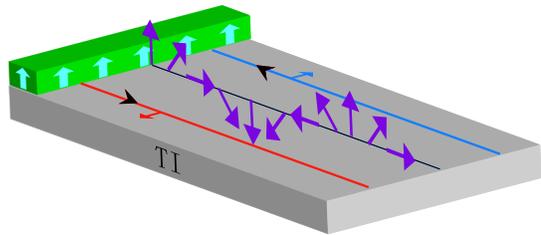}
\end{center}
\caption{(Color online) Illustration of charge and spin interferences between two
counter-propagating helical waves.
The gray block is a 3DTI with a magnetic
edge impurity (green stripe) lying in the $x$-axis on the surface.
An incident helical wave along $y$-axis with spin
polarized in the $x$-direction (blue line) is backscattered by the
magnetic edge and the spin is flipped (red line).
The interference of the two orthogonal helical waves leads to
a constant LDOS in charge channel,
but a spiral LDOS in spin channel (purple arrows) in $yz$-plane.}
\label{fig1}
\end{figure}
In the presence of TRS, the backscattering by nonmagnetic impurities
on the surface of 3DTIs is known to be forbidden ascribed to the
obtained $\pi$ Berry's phase during one full rotation of spin \cite{Ando1998,Qi2010}.
In experiments, this manifests in the invisibility of the scattering
vector ${\bf q}$ in FT-STS \cite{Yazdani2009}.
It would then be interesting to ask how the surface states
respond differently to the magnetic impurities, and what is their
characteristic signatures in the STM as well as the FT-STS measurements? With magnetic impurities,
naively we would expect the interference to be enhanced
relative to the nonmagnetic case since the backscattering is restored due to the breaking of TRS.
However, it turns out that the charge interference by magnetic
impurities is still suppressed as by the nonmagnetic impurities,
and there is still lack of notable signals in FT-STS with an ordinary STM tip.
The broken TRS would only manifest itself in the spin channel with spin-polarized STM tip \cite{Liu2009}.

To understand, we first present a simple picture of the interference between two
counter-propagating orthogonal helical waves on the surface of a 3DTI, and then
give a theoretical survey in the next section.
Suppose there is a magnetic edge impurity placed in the $x$-axis
on the surface, and a helical wave of the surface
Hamiltonian $H_0$, $\psi_1=e^{ik_Fy}(\begin{array}{cc}1&i\end{array})^{T}/\sqrt{2}$,
where the superscript ``T'' indicates the transpose, is incident
along $y$-axis with spin locked to $x$-direction, see Fig.\ref{fig1}.
This wave is then backscattered by the magnetic edge and counter-propagates in negative
$y$-direction as $\psi_2=e^{-ik_Fy}(\begin{array}{cc}1&-i\end{array})^{T}/\sqrt{2}$,
where its spin is flipped to negative $x$-direction.
A simple calculation shows that the interference between these
two counter-propagating helical waves, $\psi=\psi_1+\psi_2$, leads to a constant
charge LDOS on the surface, but a spiral spin LDOS in $yz$-plane,
$\langle\bf {s}\rangle_{\psi}=[\begin{array}{ccc}
0& \sin(2k_Fx)&\cos(2k_Fx)
\end{array}]$, where ${\bf {s}}={\mathbf\sigma}/2$ is the electron spin
operator. This interference pattern of two orthogonal helical waves in
spin channel can be detected by spin-resolved STM experiments.
To observe the standing wave of this spin interference pattern, a setup of
closed ``magnetic wall'' as shown in Fig.\ref{fig2} is proposed,
where a magnetic layer with a hollow hole is deposited on top of 3DTI surface.
Then inside this hole, a standing spin wave is formed with
$\langle s_y\rangle\sim \sin(2k_FR)/R^{1/2}$,
$\langle s_z\rangle\sim\cos(2k_FR)/R^{1/2}$, and a
loop current $j_{\phi}=-(2\hbar v_F/e)\langle s_z\rangle$ along the azimuthal direction
should be observed.
\begin{figure}[tbp]
\label{fig2}
\begin{center}
\end{center}
\caption{(Color online) Standing wave of spin interference
between two helical waves inside a closed ``magnetic wall''
on top of a 3DTI surface.
The out-of-plane spin LDOS is exhibited by the colored rings and the
in-plane spin LDOS is indicated by the dark arrows.}
\label{fig2}
\end{figure}

In fact, having in mind the transformation properties under the time-reversal that
$\Theta^{-1}\sigma^0\Theta=\sigma^{0^{T}}$, $\Theta^{-1}\sigma^a\Theta=-\sigma^{a^{T}}$,
and $\Theta^{-1}G_0(\omega,{\bf k})\Theta=G_0^{T}(\omega,{\bf
-k})$, where $\Theta=i\sigma_y$ is the time-reversal operator and $a=x,y,z$,
it can be proved in the T-matrix formulism immediately
that the spin LDOS of nonmagnetic impurities vanishes
uniformly and the charge LDOS of magnetic impurities is identical to that of
nonmagnetic impurities, by interchanging ${\bf k}$ and $-{\bf k}^{\prime}$ in the
integrals, see Eq.(\ref{ldosptdef}), using $\rm{tr}[G_0({\bf k})\sigma^aG_0({\bf
k^{\prime}})
\sigma^0]=-\rm{tr}[G_0(-{\bf
k^{\prime}})\sigma^a G_0(-{\bf k})\sigma^0]$. Therefore, in the following,
we only need to focus on the response between nonmagnetic impurity and ordinary tip
as well as that between magnetic impurity and spin-polarized tip.

\section {General formulation of Stationary phase approach to QPI on the surface of 3DTI}
\label{general formulation}
In this section, we formulate a general result of the interference features
from quasiparticle scattering on the surface of 3DTIs by nonmagnetic and
magnetic impurities with ordinary and spin-polarized tips using
the stationary phase approach \cite{Ruth1966}.

\subsection{Point impurity}
\label{pointimpurity}
We start by considering a point defect. The derivation of the LDOS
from the constant background measured in STM experiments for a
single short-range nonmagnetic or magnetic impurity with
ordinary or spin-polarized tip is given by
\begin{eqnarray}
\delta\rho_{\mu\nu}(\omega,{\bf
R})=&-&\frac{1}{\pi}\Im\int\frac{d^2kd^2k^{\prime}}{(2\pi)^4}e^{i({\bf
k-k^{\prime}})\cdot{\bf R}}\nonumber\\
&\times&\rm{tr}\left[{G}^r_0(\omega,{\bf
k})T^{\mu}(\omega,{\bf k}){G}^r_0(\omega,{\bf
k^{\prime}})\sigma^{\nu}\right].\label{ldosptdef}
\end{eqnarray}
In the above, $G^r_0(\omega,{\bf k})$ is in general the free retarded Green's function
governing the CEC under considerations, and is that of the topological surface band
in particular for our interests, where ${\bf k}=(k_x, k_y)$. $T^{\mu}=V/(1-VG^r_0(\omega))$
is the T-matrix which is ${\bf k}$-independent for a delta-potential impurity $V({\bf r})=V\delta({\bf
r})\sigma^{\mu}$, and $G^r_0(\omega)=\int \frac{d^2k}{(2\pi)^2}{G}^r_0(\omega,{\bf k})$.
The last Pauli matrix in Eq.(\ref{ldosptdef}),
$\sigma^{\nu}$, represents the polarization of the STM tip, and throughout the paper,
we use the Greek index $\mu,\nu=0$ for nonmagnetic impurity or ordinary tip, and
$\mu=a=x,y,z$ for magnetic impurity or spin-polarized tip.

Without loss of generality, we expand the T-matrix to the first order
$T^{(1)}=V\sigma^{\mu}$ and transform the integrand in Eq.(\ref{ldosptdef}) to
the diagonal basis of the topological surface bands, then the measured
LDOS becomes
\begin{eqnarray}
\delta\rho^{(1)}_{\mu\nu}(\omega,{\bf
R})=&-&\frac{V}{\pi}\Im\int\frac{d^2kd^2k^{\prime}}{(2\pi)^4}e^{i({\bf
k-k^{\prime}})\cdot{\bf
R}}\nonumber\\
&\times&\sum_{nm}\frac{\Sigma^{\mu}_{nm}({\bf k,k^{\prime}})
\Sigma^{\nu^{\ast}}_{nm}({\bf k,k^{\prime}})}
{(\omega+i\delta-\varepsilon_n)(\omega+i\delta-\varepsilon_m^{\prime})},
\label{eq2}
\end{eqnarray}
where $\varepsilon_{n,m}({\bf k})$ are the spin-splitting bands of the
surface states $|n,m{\bf k}\rangle$, and we have defined $\Sigma^{\mu}_{nm}({\bf
k,k^{\prime}})=\langle n{\bf k}|\sigma^{\mu}|m{\bf
k^{\prime}}\rangle$.
To calculate the integrals, it is tricky to notice that the
contribution to the LDOS along a given direction, say ${\bf R}=R\hat y$
(here and thereafter we shall always take the $y$-direction for example),
at large distances mainly results in the so-called ``stationary points''
${\bf k}_i$ \cite{Ruth1966}, at which
\begin{eqnarray}
\frac{\partial k_y(\varepsilon,t)}{\partial t}=
\frac{\partial k_y^{\prime}(\varepsilon^{\prime},t^{\prime})}{\partial t^{\prime}}=0,
\label{pointcnd}
\end{eqnarray}
where $t$ and $t^{\prime}$ are respectively some parameter tangent to the CEC
$\varepsilon$ and $\varepsilon^{\prime}$. This is because the phase factors
$e^{ik_y(\varepsilon,t)R}$ and $e^{ik_y^{\prime}(\varepsilon^{\prime},t^{\prime})R}$
varies rapidly with respect to $t$ and $t^{\prime}$ so that most of the
integrations cancel out except at the stationary points.
\begin{figure}[tbp]
\begin{center}
\includegraphics[width=3.8in]{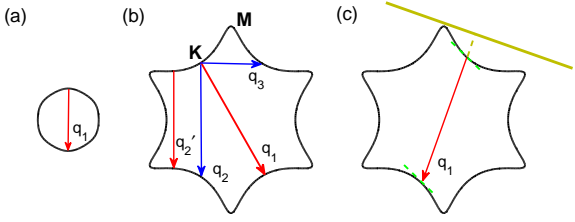}
\end{center}
\caption{(Color online) Schematic picture of CEC and stationary points
for point and edge impurities. (a) Convex CEC where
there is only one pair of stationary points connected
by the red arrow along any given direction for both point and line
impurities. (b) Concave CEC for point impurity
where there are multiple pairs of stationary points. Nonstationary
points are shown for example as blue arrows. (c)
Concave CEC for edge impurities (brown line) where the slopes (green dashed lines)
at the pair of stationary points are the same.}
\label{fig3}
\end{figure}
This condition singles out the extremal points with nonvanishing second
derivatives, such as the pairs connected by ${\bf q}_1$ in a convex CEC
in Fig.\ref{fig3} (a) and a concave CEC in
Fig.\ref{fig3} (b). Moreover, the condition (\ref{pointcnd})
also allows the turning points such as the pair connected by ${\bf q}_2^{\prime}$
in Fig.\ref{fig3}(b) where the first nonzero derivative being of the
third order. In the following, we will focus only on the extremal points for now
and leave the general discussions on the others in Sect.\ref{arbitraryCEC}.

\begin{figure}[tbp]
\begin{center}
\includegraphics[width=2.8in]{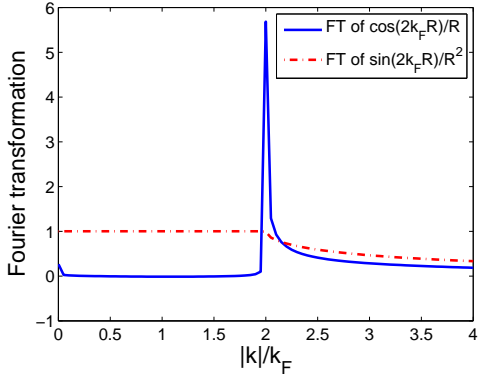}
\end{center}
\caption{(Color online) Fourier transformation of the LDOS with $R^{-1}$ and $R^{-2}$ power laws.}
\label{fig4}
\end{figure}
Having identified the pairs of stationary points on the CEC,
the integrals in Eq.(\ref{eq2}) at large distances are then been approximated
by the summation of integrals in the neighborhood of all the stationary
point pairs on the CEC. To do this, we first change the integral
variables as $d^2k=d\varepsilon dk_x/\hbar|v_{yi}|$
where $v_{yi}=\partial
\varepsilon({\bf k})/\hbar\partial k_{yi}$, and then expand the
CEC at the extremal points as $k_y=k_{yi}-(k_x-k_{xi})^2/2\rho_{xi}$,
where $\rho_{xi}=-[\partial^2k_y(\varepsilon,k_x)/\partial^2 k_{xi}]^{-1}$
is the principle radii of the curvature of the CEC at
the extremal points, which is positive for maxima while negative for minima.
Under this approximation, Eq.(\ref{eq2}) becomes
\begin{widetext}
\begin{eqnarray}
\delta\rho^{(1)}_{\mu\nu}(\omega,{\bf R})&\simeq
&-\frac{V}{\pi}\Im\sum_{mn}\sum_{ij}\int\frac{d\varepsilon_n}{(2\pi)^2}
\frac{1}{\omega+i\delta-\varepsilon_n}\frac{e^{ik_{yi}R}}{\hbar|v_{yi}|}
\int\frac{d\varepsilon_m^{\prime}}{(2\pi)^2}
\frac{1}{\omega+i\delta-\varepsilon_m^{\prime}}\frac{e^{-ik_{yj}^{\prime}R}}{\hbar|v_{yj}^{\prime}|}\nonumber\\
&\times&\int_{-\infty}^{\infty}dx\;e^{-i\frac{x^2}{2\rho_{xi}}R}
\int_{-\infty}^{\infty}dx^{\prime}\;e^{i\frac{x^{{\prime}^2}}{2\rho_{xj}^{\prime}}R}
\;\Sigma^{\mu}_{nm}({\bf
k},{\bf k}^{\prime})\Sigma^{\nu^{\ast}}_{nm}({\bf k},{\bf k}^{\prime}),
\label{eq4}
\end{eqnarray}
\end{widetext}
where $x=k_x-k_{xi}$, $x^{\prime}=k_x^{\prime}-k_{xj}^{\prime}$, and
all the quantities at the extremal points ($ij$) depend still
on the energies $\varepsilon$ and $\varepsilon^{\prime}$.
The matrix element $\Sigma^{\mu}_{nm}({\bf k}_i,{\bf k}_j^{\prime})$ is in general some
nonzero constant $C^{\mu}_{ni,mj}(\varepsilon,\varepsilon^{\prime})$,
except when the pair of stationary points are mutually time-reversal
symmetric $|n{\bf k}_i\rangle=\Theta|m {\bf k}^{\prime}_j\rangle$,
which leads to $C^0_{ni,mj}=0$ by symmetry. Examples are shown as the pairs of points
connected by ${\bf q}_1$'s in Figs.\ref{fig3}(a) and (b)
for convex and concave CEC respectively. In such a case, by letting $|n{\bf
k}\rangle=(\begin{array}{cc}e^{i\theta/2}&
e^{-i\theta/2}\end{array})^{\dagger}/\sqrt{2}$
and $|m{\bf
k}^{\prime}\rangle=(\begin{array}{cc}
e^{i\theta^{\prime}/2}&-e^{-i\theta^{\prime}/2}\end{array})^{\dagger}/\sqrt{2}$,
where $\theta=\arctan(k_y/k_x)$, we expand them around the TRP
as $\theta=\theta_i-\alpha$ and $\theta^{\prime}=\theta^{\prime}_j+\alpha^{\prime}$
where $\alpha(\alpha^{\prime})=x(x^{\prime})/k_F$, then
the matrix elements can be expanded as a series $\Sigma^0_{nm}=C^0_{ni,mj}-(x+x^{\prime})/2k_F
+[(x+x^{\prime})/2k_F]^3/3!+\cdots$.
Inserting the series into Eq.(\ref{eq4}) and integrating over the
energies, we get
\begin{widetext}
\begin{eqnarray}
\delta\rho^{(1)}_{\mu\nu}(\omega,{\bf R})&\simeq
&\frac{V}{2\pi^2\hbar^2R}\Im\sum_{mn}\sum_{ij}e^{i(k_{yi}-k_{yj}^{\prime})R}
\frac{|\rho_{xi}\rho_{xj}^{\prime}|^{\frac{1}{2}}}{|v_{yi}v_{yj}^{\prime}|}\nonumber\\
&\times&\left[e^{i(\phi_i-\phi_j)}
C^{\mu}_{ni,mj}C^{\nu^{\ast}}_{ni,mj}+\frac{1}{4k^2_FR}
\left.\left(e^{i(\phi_i+\phi_j)}\left|\rho_{xj}^{\prime}\right|
+e^{-i(\phi_i+\phi_j)}\left|\rho_{xi}\right|\right)
\right]\right|_{\varepsilon_F}
\label{eq5}
\end{eqnarray}
\end{widetext}
which is our desired result of the LDOS in STM measurements for point impurity.
In the above $\phi_i=-\frac{\pi}{4}\rm{sgn}(\rho_{xi})$,
and we have used $\int_{-\infty}^{\infty}dx e^{iCx^2}=\sqrt{\pi/|C|}e^{i\frac{\pi}{4}\rm {sgn}(C)}$
and $\int_{-\infty}^{\infty}dx
x^2e^{iCx^2}=\sqrt{\pi}/(2|C|^{3/2})e^{-i\frac{\pi}{4}\rm{sgn}(C)}$.

There are several comments regarding this result.
First of all, for a pair of non-TRS stationary points like ${\bf q}_2$
in Fig.\ref{fig3}(b), the leading power is given by the first term in Eq.(\ref{eq5}), which is of $R^{-1}$.
While for a pair of TRS stationary points as ${\bf q}_1$
in Figs.\ref{fig3} (a) and (b), the first nonvanishing contribution
to the power law is dominated
by the second term in Eq.(\ref{eq5}) as $R^{-2}$ for nonmagnetic impurity,
and for magnetic impurity but with ordinary tip.
This suppression of LDOS is a direct result of the forbiddance of backscattering
of helical waves due to TRS. Correspondingly in FT-STS measurements, there is a sharp peak at
$2k_F$ for LDOS with $R^{-1}$ power law, which is absent for $R^{-2}$ power law as
shown in Fig.\ref{fig4}. Interestingly, for magnetic impurities with
spin-polarized tip, the first term in Eq.(\ref{eq5}) dominates
no matter the pair of stationary points is TRS or not, and gives the
visibility of the TRS scattering wave vector ${\bf q}_1$ in FT-STS.
This distinct response of surface states to magnetic from that of nonmagnetic impurities
provides a crucial criteria for the breaking of TRS on the surface of TIs \cite{Xue2009}.
Second, when integrating over the energies,
we have assumed $v_{yi},v_{yj}^{\prime}\neq 0$ so that the only poles
in the complex energy plane are $\varepsilon=\varepsilon^{\prime}=\omega+i\delta$.
However, in general, it is possible that there are other poles from
$v_{yi}=0$ or $v_{yj}^{\prime}=0$, which means the stationary points in
CEC are also the extremal points in the energy-momentum dispersion. In
such a case, we shall further expand $v_{yi}$ (or $v_{yj}^{\prime}$) around $\omega$ as
$v_{yi}(\varepsilon)=v_{yi}(\omega)+(\partial
v_{yi}/\partial
\varepsilon)(\varepsilon-\omega)+\cdots$,
and take the first nonzero term to delete the singularities,
whereas this won't modify the power laws.
Finally, note that when summation over the stationary point
pairs, $(ij)$, we always choose one in the pair with
positive velocity $v_{yi}$ and
the other with negative velocity $v_{yj}^{\prime}$
to obtain the retarded Green's function.
Using the general result in Eq.(\ref{eq5}), the power laws of LDOS
for point impurity are summarized in Table. \ref{pointable}
according to the classification of TRS of stationary point
pairs on the CEC, which can well-explain the recent STM experiments observation in 3DTI.
\begin{table}[!h]
\tabcolsep 0pt \caption{Power laws from point impurity}
\vspace*{-6pt}
\begin{center}
\def\temptablewidth{0.42\textwidth}
{\rule{\temptablewidth}{1pt}}
\begin{tabular*}{\temptablewidth}{@{\extracolsep{\fill}}ccccccc}
 &  & ordinary & spin-polarized \\   \hline
nonmagnetic    &  TRP  & $R^{-2}$ & -  \\
 & non-TRP & $R^{-1}$ &  - \\
magnetic  & TRP &  $R^{-2}$  & $R^{-1}$  \\
&non-TRP & $R^{-1}$ &$R^{-1}$
       \end{tabular*}
       {\rule{\temptablewidth}{1pt}}
       \end{center}
       \label{pointable}
       \end{table}

Before going to the experiments, we first use some concrete examples
to illustrate how the formulae of (\ref{eq5}) works.
The first one is a 2D quadratic CEC, $H_Q=\hbar^2k^2/2m$, which is isotropic and
there are two degenerate spin bands, see Fig.\ref{fig5}(a).
According to our theory, the main contribution to the LDOS in this example comes
from the intraband scattering of the same spin orientation
along $y$-direction between two extremal points,
which we denote as `1' for minimum and `2' for maximum.
Then at these points explicitly we have
$k_{y2}=\rho_{x2}=k_{\varepsilon}$, $k_{y1}^{\prime}=\rho_{x1}^{\prime}=-k_{\varepsilon^{\prime}}$,
$k_{\varepsilon}=(2m\varepsilon/\hbar^2)^{1/2}$, $v_{y2}=\hbar k_{y2}/m$,
$v_{y1}^{\prime}=\hbar k_{y1}^{\prime}/m$, and $C^0_{11}=C^0_{22}=1$.
Inserting the above into Eq.(\ref{eq5}) and keeping only to the leading order,
we get $\delta\rho^{(1)}_{00}(\omega,R\hat y)\simeq-(Vm^2/\pi^2\hbar^4k_F)\cos(2k_FR)/R$,
which gives $R^{-1}$ power law. Note that the interband contribution
whereas to the LDOS is from a pair of TRS extremal points,
which has a $R^{-2}$ power law. In contrast, in the example of a 2D Dirac CEC,
$H_D=\gamma{\bf \sigma}\cdot{\bf k}$,
there is only one non-degenerate band due to the spin splitting,
see Fig.\ref{fig5}(b).
So only intraband scattering between a pair of extremal
TRP contributes such that $C^0_{ni,mj}=0$, and the leading power is expected to be $R^{-2}$.
Inserting the quantities $k_{y2}=\rho_{x2}=\varepsilon/\gamma$,
$k_{y1}^{\prime}=\rho_{x1}^{\prime}=-\varepsilon/\gamma$,
and $v_{y1(2)}=\gamma\rm{sgn}[k_{y1(2)}]/\hbar$ into Eq.(\ref{eq5}),
we get $\delta\rho^{(1)}_{00}(\omega,R\hat y)\simeq(V/4\pi^2\gamma^2)\sin(2k_FR)/R^2$,
which is the same as our expectation.
In Fig.\ref{fig4}, the Fourier transformation of the LDOS for these two examples is shown,
where we see that there is a sharp peak at $2k_F$ for LDOS of the 2DEG $H_Q$,
which is greatly broadened in the helical liquid $H_D$.

In a recent STM measurement of TI, Bi$_2$Te$_3$, with Ag-doped point
impurities\cite{Xue2009}, clear standing waves and scattering
wave vectors are imaged through
FT-STS when the Fermi surface is of hexagram shape. It is
observed that the high intensity regions are always along the
$\bar\Gamma$-$\bar M$ direction, but the intensity in $\bar\Gamma$-$\bar K$
direction vanishes. This observation can be well-understood
using our stationary phase theory. Among the three wave vectors
${\bf q}_1$, ${\bf q}_2$ (or ${\bf q}_2^{\prime}$) and ${\bf q}_3$ with high joint density
of states as shown in Fig.\ref{fig3}(b), two are connected by
the stationary points, namely ${\bf q}_1$ and ${\bf q}_2^{\prime}$,
while ${\bf q}_3$ (and ${\bf q}_2$) is not. This explains why no standing waves corresponding
to ${\bf q}_3$ are observed in FT-STS. Within the other two, stationary points
connected by ${\bf q}_1$ are also TRP which shall contribute the power law of $R^{-2}$ according
to our result, therefore its intensity in FT-STS is too weak to
observe in the experiment along $\bar\Gamma$-$\bar K$ direction.
For wave vectors ${\bf q}_2$ and  ${\bf q}_2^{\prime}$ along $\bar\Gamma$-$\bar M$ direction,
${\bf q}_2^{\prime}$ is stationary but non-TRS, out result
shows that this wave vector has $R^{-1}$ power
law, which is responsible for the high intensity reported in Ref.\cite{Xue2009}.

\subsection{Edge impurities}
\label{edgeimpurity}
Now we turn to the discussion of edge impurity. The edge impurity
is assumed to orientate in the $x$-axis on top of a 3DTI surface, see Fig.\ref{fig1},
and be modeled by the Hamiltonian
$V({\bf r})=V\delta(y)\sigma^{\mu}$. The main difference of the edge impurity
from that of the point impurity is the conservation of the momentum along
the direction of impurity edge, $k_x$, which effectively reduces one
of the integrations in Eq.(\ref{ldosptdef}),
\begin{eqnarray}
\delta\rho_{\mu\nu}(\omega,{\bf
R})=&-&\frac{1}{\pi}\Im\int\frac{d^2kd^2k^{\prime}}{(2\pi)^4}\delta_{k_x,k_x^{\prime}}e^{i({\bf
k-k^{\prime}})\cdot{\bf R}}\nonumber\\
&\times&\rm{tr}\left[{G}^r_0(\omega,{\bf
k})T^{\mu}(\omega,k_x){G}^r_0(\omega,{\bf
k^{\prime}})\sigma^{\nu}\right],\label{ldoslinedef}
\end{eqnarray}
where $T(\omega,k_x)=V(1-V G^r_0(\omega,k_x))^{-1}$ and
$G^r_0(\omega,k_x)=\int \frac{dk_y}{2\pi} G^r_0(\omega,{\bf k})$.
In the presence of edge impurity, we are usually interested in the
LDOS along the direction perpendicular to the impurity edge.
Then the main contribution to the LDOS now comes from such pairs of
stationary points where their momentum transfer
${\bf q}={\bf k}_i-{\bf k}_j^{\prime}$ is normal to the impurity edge
and the ``slopes'' at the pair of points are the same on the CEC,
\begin{eqnarray}
\frac{\partial}{\partial t}\left[k_y(\varepsilon,t)
-k_y^{\prime}(\varepsilon^{\prime},t)\right]=0. \label{linecnd}
\end{eqnarray}
This condition allows more possibilities than those implied
by Eq.(\ref{pointcnd}) for point impurity.
One such example is shown schematically as ${\bf q}_1$ in Fig.\ref{fig3}(c)
where the pair of stationary points has the same slope
$\alpha_i=\alpha_j^{\prime}$ (the green dashed lines),
with $\alpha_{i}=\partial k_y(\varepsilon,k_{xi})/\partial
k_{xi}$, which is
not necessary to be zero. Following the same logic as the discussions
in point impurity case in the last subsection, the CEC is expanded around the stationary points as
$k_y=k_{yi}+\alpha_{i}(k_x-k_{xi})-(k_x-k_{xi})^2/2\rho_{xi}$,
and the LDOS is approximated by
\begin{widetext}
\begin{eqnarray}
\delta\rho^{(1)}_{\mu\nu}(\omega,{\bf R})\simeq
&&-\frac{V}{\pi}\Im\sum_{mn}\sum_{ij}\int\frac{d\varepsilon_n}{(2\pi)^2}
\frac{1}{\omega+i\delta-\varepsilon_n}\frac{e^{ik_{yi}R}}{\hbar|v_{yi}|}
\int\frac{d\varepsilon_m^{\prime}}{(2\pi)^2}
\frac{1}{\omega+i\delta-\varepsilon_m^{\prime}}\frac{e^{-ik_{yj}^{\prime}R}}{\hbar|v_{yj}^{\prime}|}\nonumber\\
&&\times\int_{-\infty}^{\infty}dxe^{-i\frac{x^2}{2\rho_{xi}}R}
\int_{-\infty}^{\infty}dx^{\prime}e^{i\frac{x^{{\prime}^2}}{2\rho_{xj}^{\prime}}R}
e^{i\alpha_{xi}(x-x^{\prime})}\delta_{x,x^{\prime}}
\left[C^{\mu}_{ni,mj}C^{\nu^{\ast}}_{ni,mj}+\frac{(x+x^{\prime})^2}{4k_F^2}\right].
\label{eq8}
\end{eqnarray}
\end{widetext}
Note that though the requirement of momentum conservation,
$\alpha_{xi}(x-x^{\prime})=0$, makes Eq.(\ref{eq8}) the same as
that in point impurity case except for a factor $\delta_{x,x^{\prime}}$,
the definition of stationary points are physically quite different
and includes more terms in the summation of stationary point pairs $(ij)$.
After integrating out $k_x$ and the energies, it leads to the
final result of LDOS in the presence of edge impurity
\begin{widetext}
\begin{eqnarray}
\delta\rho^{(1)}_{\mu\nu}(\omega,{\bf R})\simeq
\frac{V}{(2\pi)^2\hbar^2}\sqrt{\frac{2}{\pi
R}}\Im\sum_{mn}\sum_{ij}\left(\frac{|P_{ij}|^{1/2}e^{i(k_{yi}-k_{yj}^{\prime})R}}{|v_{yi}v_{yj}^{\prime}|}
\left[C^{\mu}_{ni,mj}C^{\nu^{\ast}}_{ni,mj}e^{i\Phi_{ij}}+\frac{|P_{ij}|}{k^2_FR}e^{-i\Phi_{ij}}\right]
\right),
\label{eq9}
\end{eqnarray}
\end{widetext}
where $P_{ij}=\rho_{xi}\rho_{xj}^{\prime}/(\rho_{xj}^{\prime}-\rho_{xi})$
and $\Phi_{ij}=-\frac{\pi}{4}\text{sgn}(P_{ij})$. In the above we have assumed
$\rho_{xi}\neq \rho_{xj}^{\prime}$, that is we are not considering the case where the CEC
near the pair of stationary points is nested,
otherwise the quadratic terms in the expansion of CEC
cancel out exactly and higher orders expansion should be employed.
The power laws of LDOS from edge impurity are summarized in Talbe. \ref{linetable},
which shall be used to explain the recent STM measurements
for edge impurities\cite{Manoharan2009,Alpichshev2009}.
\begin{table}[!h]
\tabcolsep 0pt \caption{Power laws from edge impurity} \vspace*{-6pt}
\begin{center}
\def\temptablewidth{0.42\textwidth}
{\rule{\temptablewidth}{1pt}}
\begin{tabular*}{\temptablewidth}{@{\extracolsep{\fill}}ccccccc}
 &  & ordinary & spin polarized \\   \hline
nonmagnetic  &  TRP  & $R^{-3/2}$ & -  \\
 & non-TRP & $R^{-1/2}$ &  - \\
magnetic & TRP &  $R^{-3/2}$  & $R^{-1/2}$  \\
&non-TRP & $R^{-1/2}$ &$R^{-1/2}$
       \end{tabular*}
       {\rule{\temptablewidth}{1pt}}
       \end{center}
       \label{linetable}
       \end{table}

To have a feeling of how Eq.(\ref{eq9}) works, again we apply it first to
the examples of $H_Q$ and $H_D$ used in point impurity.
A few lines of calculations yield that
for 2D quadratic dispersion,
$\delta\rho_{00}^{(1)}(\omega,R\hat{y})=(Vm^2/2\pi^2\hbar^4k_F^{3/2})
\sin(2k_FR-\frac{\pi}{4})/\sqrt{\pi R}$, which recovers the experimental observation
in a 2DEG \cite{Crommie1993}. While for 2D Dirac dispersion,
$\delta\rho_{00}^{(1)}(\omega,R\hat{y})=(V/8\pi^2\gamma^2\sqrt{\pi k_F})
\sin(2k_FR+\frac{\pi}{4})/R^{3/2}$, which is a result of the absence of backscattering
in helical liquid. Similar information in reciprocal space can be extracted via FT-STS
as exhibited in Fig.\ref{fig4}, where a notable sharp peak appears at $2k_F$ for a 2DEG,
which disappears contrastively for helical liquid.
\begin{figure}[tbp]
\begin{center}
\includegraphics[width=3.2in]{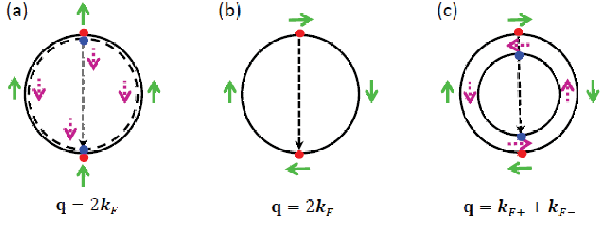}
\end{center}
\caption{(Color online) Schematic CEC of (a) quadratic, (b) Dirac,
and (c) Rashba dispersions. The spin orientations for each degenerate band
are indicated respectively by the green (solid) and purple (dotted)
arrows. The stationary points are represented by red and blue dots,
which are connected by the scattering vector ${\bf q}$ shown as dashed arrows.
The intraband scattering comes from the stationary points with the same color,
while the interband scattering comes from those with different colors.}
\label{fig5}
\end{figure}

In the recent experiment by Gomes {\it et al.},
a nonmagnetic step is imaged by STM topography in clean Sb $(1 1 1)$ surface with
nontrivial topology\cite{Manoharan2009}. The Fermi surface consists
of one electron pocket at $\bar\Gamma$, which is surrounded by six hole pockets
in $\bar\Gamma$-$\bar M$ direction, where the surface dispersion
shows a Rashba spin splitting. The measured LDOS in $\bar\Gamma$-$\bar M$ direction is
fitted by a single $q$-parameter using
the zeroth-order of Bessel function of the first kind,
see Fig.2(c) in Ref.\cite{Manoharan2009}, which agrees exactly
with our prediction in Table. \ref{linetable}. Along $\bar\Gamma$-$\bar M$ direction, the surface
band can be modeled by a Rashba Hamiltonian where the LDOS is dominated
by interband scattering between a pair of non-TRS stationary points,
see Fig.\ref{fig5}(c).
According to our analysis, the Friedel oscillation has $R^{-1/2}$
power law, which is the asymptotic expansion of $J_0(k_FR)$ at large distances.
Another STM experiment studies the edge impurity, however in Bi$_2$Te$_2$, is
the work by Alpichshev {\it et. al.} \cite{Alpichshev2009},
where a nonmagnetic step defect with height about one unit cell is shaped on
crystal surface. A strongly damped oscillation
is reported when the bias voltage is at the energy with
a convex Fermi surface as shown in Fig.\ref{fig3}(a). Though no
fitting of the experimental data is estimated in this region, our
results predict a $R^{-3/2}$ power law. Pronounced oscillations
at higher bias voltages where the hexagon warping effect emerges
are observed with $R^{-1}$ fitting. Despite of the quantitative difference
with our result of $R^{-1/2}$, this $R^{-1}$ oscillations have been
explained in several other works \cite{Zhou2009,Lee2009} beyond
our simple model.

Finally, it is emphasized that the results in Tables. \ref{pointable} and
\ref{linetable} provide a quantitative description of the QPI by
magnetic impurities in general, and the interference between
two orthogonal helical waves as discussed in Sect. \ref{standingwave} in particular.
The interference of helical waves corresponds to the
scattering between two TRS stationary points, like the ${\bf q}_1$'s in
Figs. \ref{fig3}(a), (b) and (c). The interesting thing is that
the LDOS in charge and spin channels from the very same pair of
TRS stationary points has quite distinct behaviors. With magnetic impurities,
the power laws of charge LDOS are $R^{-2}$ and $R^{-3/2}$
for point and edge impurities respectively, which is a result of TRS and
have higher power
indices than the $R^{-1}$ and $R^{-1/2}$ modulations of the spin-polarized LDOS,
which manifests the TRS breaking.
So that the charge LDOS decays much faster than the spin LDOS, and
to tell the different response of topological surface states to
magnetic impurities from that of the nonmagnetic ones \cite{Xue2009},
spin-resolved STM experiments in the proposed closed ``magnetic wall''
structure in Fig.\ref{fig2} are called on in particular.

\subsection{Friedel oscillations in arbitrarily-shaped CEC}
\label{arbitraryCEC}
In this section, we shall complete the most general formulation of the QPI
on the surface of 3DTI by inclusion those special cases which
have been skipped over in the previously discussions.

For point
impurity, under the condition (\ref{pointcnd}), we have made the
extremal points assumption in Sect. \ref{pointimpurity}
that the expansion of CEC around the stationary points has nonvanishing
second derivatives. However it is in general possible that $\rho_{xi}$
diverges so that we need to go to the third or even higher order
expansions. For example, when the stationary points are also
turning points on the CEC. While for edge impurity, under the condition (\ref{linecnd}),
besides the above example of turning points where
$\alpha_{i}=\alpha_{j}^{\prime}=\rho_{xi}^{-1}=\rho_{xj}^{{-1}^{\prime}}=0$,
it is also possible that $\rho_{xi}=\rho_{xj}^{\prime}\neq 0$ but
$P_{ij}$ diverges. This happens when the CEC near the stationary points
is nested, and we need to go beyond quadratic
term expansions till some power at which the exact nesting property
is unbalanced.
\begin{table}[!h]
\tabcolsep 0pt \caption{General results of power laws for point impurity}
\vspace*{-6pt}
\begin{center}
\def\temptablewidth{0.42\textwidth}
{\rule{\temptablewidth}{1pt}}
\begin{tabular*}{\temptablewidth}{@{\extracolsep{\fill}}ccccccc}
 &  & ordinary & spin-polarized \\   \hline
nonmagnetic&  TRP  &
$R^{-(\frac{1}{l}+\frac{1}{h})-\frac{2}{\rm{min}(l,h)}}$
& -  \\
 & non-TRP & $R^{-(\frac{1}{l}+\frac{1}{h})}$ &  - \\
magnetic & TRP &
$R^{-(\frac{1}{l}+\frac{1}{h})-\frac{2}{\rm{min}(l,h)}}$
  & $R^{-(\frac{1}{l}+\frac{1}{h})}$  \\
&non-TRP & $R^{-(\frac{1}{l}+\frac{1}{h})}$
&$R^{-(\frac{1}{l}+\frac{1}{h})}$
       \end{tabular*}
       {\rule{\temptablewidth}{1pt}}
       \end{center}
       \label{pointgeneral}
       \end{table}

To understand the LDOS behavior in STM experiments in these situations, we assume
that the first nonvanishing terms in the expansions of $k_y$ and $k_y^{\prime}$ around the
stationary points are in general respectively $k_y=k_{yi}+\beta^{(l)}_{i}(k_x-k_{xi})^l$
and $k_y^{\prime}=k_{yj}^{\prime}+\beta_{j}^{\prime^{(h)}}(k_x^{\prime}-k_{xj}^{\prime})^h$,
where $l,h\in Z$ and the $\beta$'s are the expansion
coefficients which is explicitly $\beta^{(l)}_{i}=(\partial^l k_y/\partial k_{xi}^l)/l!$
and similar for $\beta_{j}^{\prime^{(h)}}$.
Notice that for edge impurity, if $l=h$ one more constrain $\beta^{(l)}_{i}\neq
\beta_{j}^{\prime^{(h)}}$ is further required in particular.
Then similar calculations as performed in Sects. \ref{pointimpurity} and
\ref{edgeimpurity} lead to the following results
for point and edge impurities separately, that
\begin{widetext}
\begin{eqnarray}
\rho^{(1)}_{\mu\nu}(\omega,{\bf R})&\propto&
\frac{V}{R^{\frac{1}{l}+\frac{1}{h}}}\text{Im}\sum_{mn}\sum_{ij}\left\{\frac{e^{i(k_{yi}-k_{yj}^{\prime})R}}
{|v_{yi}v_{yj}^{\prime}||\beta^{(l)}_{xi}|^{\frac{1}{l}}|\beta^{\prime^{(h)}}_{xj}|^{\frac{1}{h}}}
\left[C^{\mu}_{ni,mj}C^{\nu^{\ast}}_{ni,mj}
+\frac{1}{4k^2_F}\left(\frac{1}{|\beta^{(l)}_{xi}|^{\frac{2}{l}}R^{\frac{2}{l}}}
+\frac{1}{|\beta_{xj}^{\prime^{(h)}}|^{\frac{2}{h}}R^{\frac{2}{h}}}
\right)\right]\right\}_{\varepsilon_F}\label{eq10}\\
\rho^{(1)}_{\mu\nu}(\omega,{\bf R})
&\propto&\frac{V}{R^{\frac{1}{\text{max}(l,h)}}}\Im\sum_{mn}\sum_{ij}
\left\{\frac{e^{i(k_{yi}-k_{yj}^{\prime})R}}
{|v_{yi}v_{yj}^{\prime}||\beta^{(l)}_{xi}-\beta_{xj}^{\prime^{(h)}}|^{\frac{1}{\text{max}(l,h)}}}
\left[C^{\mu}_{ni,mj}C^{\nu^{\ast}}_{ni,mj}
+\frac{1}{k^2_F}\frac{1}{(R|\beta^{(l)}_{xi}-\beta_{xj}^{\prime^{(h)}}|)^
{\frac{2}{\text{max}(l,h)}}}\right]\right\}
_{\varepsilon_F}\label{eq11}
\end{eqnarray}
\end{widetext}
where the notations $\rm{min}(l,h)$ and $\rm{max}(l,h)$
represent taking the minimum or the maximum one in-between $l$ and $h$.
The corresponding power laws of LDOS are listed in Tables. \ref{pointgeneral} and \ref{linegenral}.
\begin{table}[!h]
\tabcolsep 0pt \caption{General results of power laws for edge impurity}
\vspace*{-6pt}
\begin{center}
\def\temptablewidth{0.42\textwidth}
{\rule{\temptablewidth}{1pt}}
\begin{tabular*}{\temptablewidth}{@{\extracolsep{\fill}}ccccccc}
 &  & ordinary & spin-polarized \\   \hline
nonmagnetic   & TRP  & $R^{-\frac{3}{\text{max}(l,h)}}$ & -  \\
 & non-TRP & $R^{-\frac{1}{\text{max}(l,h)}}$ &  - \\
magnetic& TRP &  $R^{-\frac{3}{\text{max}(l,h)}}$  & $R^{-\frac{1}{\text{max}(l,h)}}$  \\
&non-TRP & $R^{-\frac{1}{\text{max}(l,h)}}$
&$R^{-\frac{1}{\text{max}(l,h)}}$
       \end{tabular*}
       {\rule{\temptablewidth}{1pt}}
       \end{center}
       \label{linegenral}
       \end{table}
We see that by setting $l=h=2$, Tables. \ref{pointable} and \ref{linetable} are recovered.
Eqs. (\ref{eq10}) and (\ref{eq11}) in principle can be used to
describe the QPI on an arbitrarily-shaped CEC, which is very useful
as a first-glance guide of the freshly cooked experimental data.

\section{Conclusions}
\label{conclusion}
In conclusion, general analytic expressions of the LDOS are derived
in Eqs.(\ref{eq5}), (\ref{eq9}), (\ref{eq10}), and (\ref{eq11})
using the stationary phase approach,
for nonmagnetic and magnetic point and edge impurities
in ordinary and spin-polarized STM experiments.
The power laws of Friedel oscillation are extracted in
Tables. \ref{pointable} to \ref{linegenral} in particular.
The QPI from magnetic impurities are focused on, in which the interference
of charge intensity is indistinguishable from that of nonmagnetic impurities,
while the spin intensity of magnetic impurities shows distinctive FT-STS patterns,
which is proposed to be realized in a closed ``magnetic wall'' setup through
spin-polarized STM measurements.
Our results depend only on the TRS as well as the local geometry around the
stationary points on the CEC, so that they are suitable especially as a
first-glance guide of the experiments with ever changing Fermi surfaces,
like the emergence of the
hexagon warping effects in Bi$_2$Te$_3$.

\acknowledgements
The author would like to thank Xiao-Liang Qi and Shou-Cheng Zhang
for illuminating discussions.
This work is supported by the NSFC (Grant Nos.
11004212, 10704080, 60877067, and 60938004) and the
STCSM (Grant Nos. 08dj1400303 and 11ZR1443800).

\end{document}